\documentclass[aps,prb,twocolumn,superscriptaddress,showpacs,floatfix,amsmath,amssymb]{revtex4-1}

\usepackage{graphicx}
\usepackage{dcolumn}
\usepackage{bm}
\usepackage[below]{placeins}
\usepackage{setspace} 

\begin{document}


\title {Doping influence of spin dynamics and magnetoelectric effect in hexagonal Y$_{0.7}$Lu$_{0.3}$MnO$_{3}$}



\author{W. Tian}
\email{wt6@ornl.gov}
\affiliation{Quantum Condensed Matter Division,
Oak Ridge National Laboratory, Oak Ridge, Tennessee 37831, USA }

\author{Guotai Tan}
\email{tanbj2008@gmail.com}
\affiliation{Department of Physics,
Beijing Normal University, Beijing 100875, China}

\author{Liu Liu}
\affiliation{Department of Physics, Beijing Normal University,
Beijing 100875, China}

\author{Jinxing Zhang}
\affiliation{Department of Physics, Beijing Normal University,
Beijing 100875, China}

\author{Barry Winn}
\affiliation{Quantum Condensed Matter Division, Oak Ridge National
Laboratory, Oak Ridge, Tennessee 37831, USA }

\author{Tao Hong}
\affiliation{Quantum Condensed Matter Division, Oak Ridge National
Laboratory, Oak Ridge, Tennessee 37831, USA }

\author{J. A. Fernandez-Baca}
\affiliation{Quantum Condensed Matter Division, Oak Ridge National
Laboratory, Oak Ridge, Tennessee 37831, USA }
\affiliation{Department of Physics and Astronomy, The University of
Tennessee, Knoxville, Tennessee 37996, USA}

\author{Chenglin Zhang}
\affiliation{Department of Physics and Astronomy, The University of
Tennessee, Knoxville, Tennessee 37996, USA}
\affiliation{Department of Physics and Astronomy, Rice University, Houston, Texas 77005, USA}

\author{Pengcheng Dai}
\affiliation{Department of Physics and Astronomy, The University of
Tennessee, Knoxville, Tennessee 37996, USA}
\affiliation{Department of Physics and Astronomy, Rice University, Houston, Texas 77005, USA}

\date{\today}

\begin{abstract}
We use inelastic neutron scattering to study spin waves and their
correlation with the magnetoelectric effect in
Y$_{0.7}$Lu$_{0.3}$MnO$_3$. In the undoped YMnO$_3$ and LuMnO$_3$,
the Mn trimerization distortion has been suggested to play a key
role in determining the magnetic structure and the magnetoelectric
effect. In Y$_{0.7}$Lu$_{0.3}$MnO$_3$, we find a much smaller
in-plane (hexagonal $ab$-plane) single ion anisotropy gap that
coincides with a weaker in-plane dielectric anomaly at $T_N$. Since
both the smaller in-plane anisotropy gap and the weaker in-plane
dielectric anomaly are coupled to a weaker Mn trimerization
distortion in Y$_{0.7}$Lu$_{0.3}$MnO$_3$ comparing to YMnO$_3$ and
LuMnO$_3$, we conclude that the Mn trimerization is responsible for
the magnetoelectric effect and multiferroic phenomenon in
Y$_{1-y}$Lu$_{y}$MnO$_{3}$.
\end{abstract}


\maketitle

Driven by modern technology towards device miniaturization, there is
considerable interest in multiferroic materials which exhibit both
magnetic order and electrical polarization
\cite{Fiebig-JPD-2005,Fiebig-science-2005,Tokura-science-2006,Eerenstein-nature-2006,
Kimura-nature-2003, Hur-nature-2004,cheong-nature-mater-2007}. The
hexagonal manganite $R$MnO$_3$\cite{Fiebig-natute-2002,
Lottermoser-nature-2004}(where $R$ is a rare-earth element with
relatively small ionic radius) is a prototypical example of the
so-called type-I multiferroics \cite{Harris-book} with ferroelectric
order at $T_c$$\sim$900 K \cite{Choi-Nature-mater-2010} and an
antiferromagnetic (AFM) order at much lower temperature,
$T_N\sim$100 K \cite{Tomuta-2001}. A large dielectric anomaly occurs
at T$_N$ \cite{Iwata-1998,Huang-1997,Katsufuji-2001} indicating
strong magnetoelectric (ME) coupling in these materials. There has
been a large amount of experimental work in recent years, with the
aim to understand the microscopic mechanism for the coupling between
magnetic and electric degrees of freedom in these materials.
Although it is generally believed that the spin-lattice coupling
plays an important role in determining the complex properties in
$R$MnO$_3$, much is unclear concerning the factors that influences
the magnitude of the ME coupling
\cite{Lee-Nature-2008,Aken-LuMnO3-PRB-2004}. In this Letter, we use
inelastic neutron scattering (INS) and dielectric constant
measurements to show that the magnitude of the ME coupling in
multiferroic Y$_{0.7}$Lu$_{0.3}$MnO$_3$ is directly coupled to the
strength of the Mn trimerization distortion in these materials in
the AFM phase. Our results thus provide direct evidence that the Mn
trimerization is responsible for the ME effect and multiferroic
phenomenon in Y$_{1-y}$Lu$_{y}$MnO$_{3}$.

\begin{figure} [htp!]
\centering\includegraphics[width=2.6in]{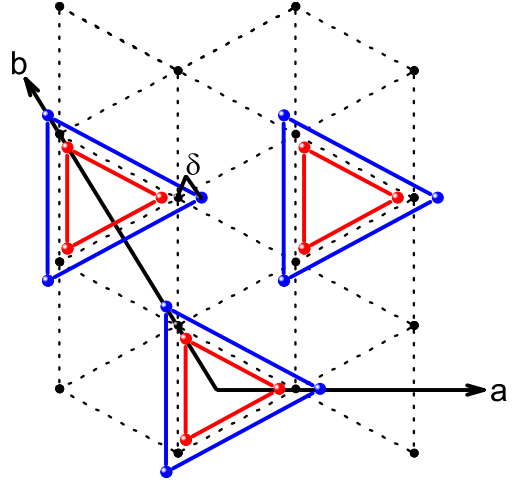}
\caption{\label{fig:trimerization}(Color online) Schematic drawing
of the Mn trimerization distortion below T$_N$. The dotted line
depicts the ideal triangular lattice in the hexagonal $ab$ plane
with the Mn (filled black circle) $x$ position at $x_c$ = 1/3. The
blue and red balls represent the Mn position $x$ in the $z$ = 0
plane for YMnO$_3$ and LuMnO$_3$ illustrating opposite distortion
directions of the Mn trimers, expansion in YMnO$_3$ and contraction
in LuMnO$_3$, respectively. $\delta$ = $\mid x-x_c\mid$ is the
trimerization distortion parameter as described in the text.}
\end{figure}

The undoped compounds YMnO$_3$ and LuMnO$_3$ are characteristic
hexagonal manganites where the Mn$^{3+}$ ions (Mn $x$ position at
$x$$\sim$1/3) form a nearly ideal triangular lattice in the
$ab$-plane above $T_N$. YMnO$_3$ and LuMnO$_3$ undergo AFM
transitions (to two different magnetic structures) at $T_N$$\sim$75
K and $T_N$$\sim$88 K respectively, accompanied by an isostructural
transition with large atomic displacement for all atoms in the unit
cell. In particular, a distinct change of the Mn atomic position,
namely the Mn trimerization distortion, occurs in the basal plane at
$T_N$ \cite{Lee-Nature-2008}. As illustrated in
Fig.~\ref{fig:trimerization}, the Mn trimers distort in opposite
directions in YMnO$_3$ and LuMnO$_3$, expanding for YMnO$_3$ and
contracting for LuMnO$_3$. A recent theoretical study finds that the
different magnetic structures of YMnO$_3$ and LuMnO$_3$ are
determined by the different trimerization directions in these
compounds \cite{Solovyev-PRB-2012-theory}. Moreover, the dielectric
anomaly at $T_N$ is observed only in $\varepsilon_{ab}$ but not in
$\varepsilon_{c}$ for both YMnO$_3$ and LuMnO$_3$
\cite{Katsufuji-2001}.  Although these studies suggest that the Mn
trimerization may play a key role in determining the magnetic
structure and the ME effects in Y$_{1-y}$Lu$_y$MnO$_3$, there are no
experimental studies to determine the connection between the Mn
trimerization and ME coupling. Y$_{1-y}$Lu$_y$MnO$_3$ is an ideal
system for such a study due to the following reasons: (1) since both
Y and Lu are nonmagnetic, Y$_{1-y}$Lu$_y$MnO$_3$ is a clean system
to study the magnetism of the Mn triangular lattice and its
correlation with the ME effects; (2) the strength of the Mn
trimerization distortion can be tuned in Y$_{1-y}$Lu$_y$MnO$_3$.
With increasing Lu concentration, the Mn atomic position $x$ changes
from 0.340 for YMnO$_3$, larger than $x_c$=1/3 for an ideal
triangular lattice, to 0.331 for LuMnO$_3$, smaller than 1/3.  At
$y$$\sim$0.3 Lu doping, the Mn atomic position $x$ crosses the
critical value $x_c$=1/3 with a perfect triangular lattice without
trimerization distortion \cite{Park-PRB-2010}.

All measurements reported here were performed on single crystal
samples. Large Y$_{1-y}$Lu$_{y}$MnO$_3$ single crystals with nominal
value of $y$=0.3 were grown by the floating zone method under 4
atmospheres of oxygen flow. The crystals cut from the long rods were
then annealed at 1350$^{\circ}$C for 24 hours in a flowing argon
atmosphere. For the magnetic susceptibility and dielectric constant
measurements, the single crystal was cut into thin plates with
$ab$-axes lying in the plane and $c$-axis pointing out of the plane.
The magnetic susceptibility was measured using a Quantum Design
Magnetic Properties Measurement System with magnetic field applied
along the $c$-axis. The dielectric constant was measured using a LRC
meter with electric field applied perpendicular and parallel to the
$c$-axis and data were taken at 3.5 V ac driving voltage and 100 kHz
frequency.

\begin{figure} [!htp]
\centering\includegraphics[width=3in]{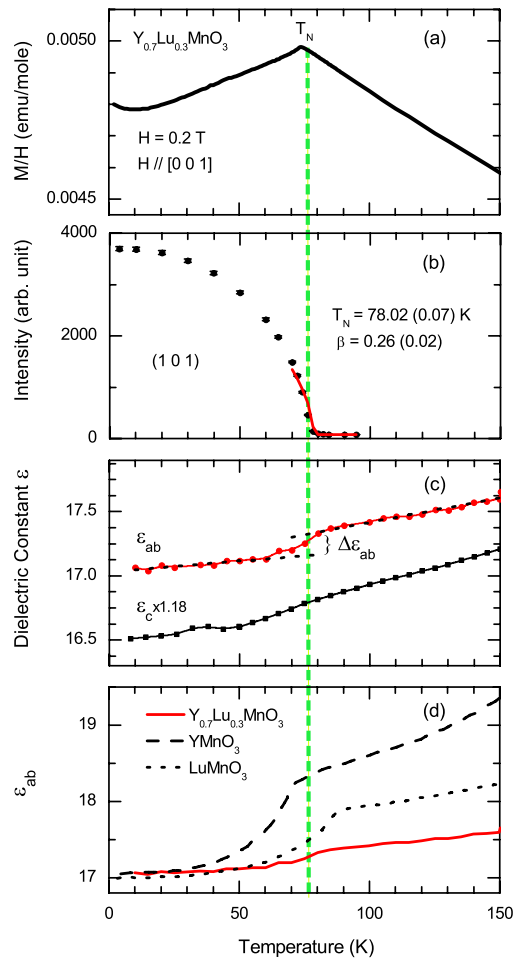}
\caption{\label{fig:tn}(Color online) Magnetic susceptibility,
magnetic order parameter and dielectric constant measurements of
Y$_{0.7}$Lu$_{0.3}$MnO$_3$. The dashed green line depicts the Neel
temperature $T_N$. (a) Magnetic susceptibility versus temperature
measured with applied magnetic field parallel to the $c$-axis. (b)
Integrated intensity of the (1 0 1) magnetic Bragg reflection as a
function of temperature. The solid red line is a fit of the data to
the power law as described in the text. (c) In-plane and
out-of-plane dielectric constant measured with the electric field
applied perpendicular and parallel to the $c$-axis. $
\Delta$$\varepsilon_{ab}$ represents the critical in-plane
dielectric constant change at $T_N$. (d) Comparison of the in-plane
dielectric constant $\varepsilon_{ab}$ between YMnO$_3$, LuMnO$_3$
and Y$_{0.7}$Lu$_{0.3}$MnO$_3$. The data for YMnO$_3$
($\varepsilon_{ab}$-2.1) and LuMnO$_3$ ($\varepsilon_{ab}$+0.3) are
from Ref. \onlinecite{Katsufuji-2001} and are plotted with an offset
of -2.1 and 0.3 for YMnO$_3$ and LuMnO$_3$, respectively.}
\end{figure}

\begin{figure} [htp!]
\centering\includegraphics[width=2.8in]{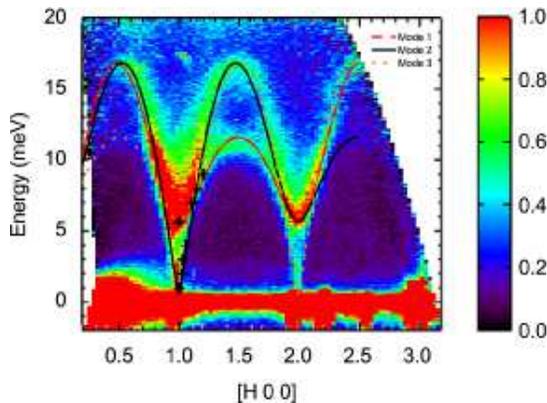}
\caption{\label{fig:spinwave}(Color online) INS spectra of
Y$_{0.7}$Lu$_{0.3}$MnO$_3$ along ($H$ 0 0) from the HYSPEC
measurements at 4 K. The crossed symbols are data points obtained
from TAS measurements. The lines are the calculated dispersion using
the fitting parameters as described in the text.}
\end{figure}

A single crystal with a mass of $\sim$4 gram was used for the
neutron scattering experiments. The crystal was mounted on a
aluminum plate and oriented in the ($H$ 0 $L$) scattering plane. The
sample was then sealed in aluminum sample can under helium
atmosphere and cooled using a closed-cycle He refrigerator. The
neutron experiments were carried out using the HB-1A and CTAX
triple-axis spectrometers (TAS) located at the High Flux Isotope
Reactor (HFIR), and the Hybrid Spectrometer (HYSPEC) located at the
Spallation Neutron Source (SNS) at Oak Ridge National Laboratory.
HB-1A is a fixed incident energy TAS (E$_i$=14.64 meV), and CTAX is
a cold neutron TAS. Collimations of $48'$-$48'$-sample-$40'$-$80'$
downstream from the reactor to the detector was used for the HB-1A
experiment with two pyrolitic graphite (PG) filters placed before
the sample to eliminate higher-order contaminations in the beam. The
CTAX experiment was performed with a fixed final energy of E$_f$=3
meV (energy resolution is $\sim$0.1 meV FWHM at elastic condition)
and collimations of guide-open-sample-$80'$-open. Higher-order
contaminations were removed by a cooled Be filter placed between the
sample and the analyzer. The HYSPEC experiment was carried out using
an incident energy of E$_i$=25 meV with a Fermi chopper spinning at
420 Hz.

The Y$_{0.7}$Lu$_{0.3}$MnO$_3$ sample was characterized by the
magnetic susceptibility and neutron scattering magnetic order
parameter measurements. Fig.~\ref{fig:tn} (a) shows the magnetic
susceptibility measured with $H \parallel c$ exhibiting a kink at
$\sim78$ K indicating the AFM transition. The order parameter
plotted in Fig.~\ref{fig:tn} (b) was measured by monitoring the
strong magnetic Bragg peak (1 0 1) as a function of temperature. The
integrated intensity was obtained by fitting the (1 0 1) rocking
curve measured at each temperature to a Gaussian function with a
constant background. As illustrated by the dashed green line, the
AFM transition at $T_N\sim78$ K was observed in both measurements
consistent with previous reports \cite{Park-PRB-2010}. The solid red
line in Fig.~\ref{fig:tn} (b) is a fit to a power-law
\textit{I(T)=I$_0$[(T$_N$-T)/T$_N$)]$^{2\beta}$} that yields
$T_N$$\approx$78.02$\pm$0.07 K and $\beta$$\approx$0.26$\pm$0.02,
where $\beta$ is the critical exponent. The yielded $\beta$ value
$\sim$0.26 is between the theoretical values of a 2D ($\beta$=0.125)
and a 3D ($\beta$=0.326) Ising system in good agreement with a prior
study \cite{YMnO3-INS1-cryst}.

The spin dynamics of the Mn$^{3+}$ ions has been investigated in
detail in YMnO$_3$, LuMnO$_3$, and HoMnO$_3$
\cite{YMnO3-INS1-cryst,Vajk-PRL-2005,
Lewtas-PRB-2010,YMnO3-INS1-powder,YMnO3-INS2-powder}.
Fig.~\ref{fig:spinwave} shows the INS spectra of
Y$_{0.7}$Lu$_{0.3}$MnO$_3$ at 4K projected along the ($H$ 0 0)
direction. The measured spectra is very similar to that reported for
YMnO$_3$ and LuMnO$_3$. At (1 0 0), two anisotropy gaps $\Delta_{1}$
and $\Delta_{2}$ were observed for both YMnO$_3$ and LuMnO$_3$,
$\Delta_{1}$ $\approx$ 2.4 meV and $\Delta_{2}$ $\approx$ 5.4 meV
for YMnO$_3$ \cite{YMnO3-INS1-cryst}, and $\Delta_{1}$ $\approx$ 2.5
meV and $\Delta_{2}$ $\approx$ 6.5 meV for LuMnO$_3$
\cite{Lewtas-PRB-2010}, respectively. Compared to the parent
compounds, our measurements show a much smaller $\Delta_{1}$ gap in
Y$_{0.7}$Lu$_{0.3}$MnO$_3$. The same sample was then measured using
both cold and thermal TAS at HFIR to better characterize the two
gaps in Y$_{0.7}$Lu$_{0.3}$MnO$_3$.

\begin{table*}[hpt!]
\caption{\label{tb:comp} Compare the lattice constants (space group
 $P6_{3}cm$), $T_N$, trimerization distortion parameter $\delta$, exchange
constant J, single ion anisotropy parameters $D_y$, $D_z$, and the
critical dielectric constant change $\Delta$$\varepsilon_{ab}$
between YMnO$_3$, LuMnO$_3$ and Y$_{0.7}$Lu$_{0.3}$MnO$_3$.}
\begin{ruledtabular}
\begin{tabular}{p{62pt}cccccccc}
\centering{} & Lattice (\AA) & T$_N$ (K) & $ \delta$ &J (meV) & D$_y$ (meV) & D$_z$ (meV)& $ \Delta$$\varepsilon_{ab}$ & Refs.\\
\hline
\centering {YMnO$_{3}$} & $a$=6.132, $c$=11.452 & 75 &  0.007 &2.4 & 0.033 & 0.32 & 1.02 & Ref.\onlinecite{YMnO3-INS1-cryst}\\
\centering {Y$_{0.7}$Lu$_{0.3}$MnO$_{3}$} &$a$=6.103(2), $c$=11.403(1)  & 78 & 0.001 & 2.57(5) & 0.0017(2) & 0.35(1) & 0.17 & This Work\\
\centering {LuMnO$_{3}$} & $a$=6.05, $c$=11.4  & 88 & 0.003 & 2.9 & 0.035 & 0.4 & 0.77 & Ref.\onlinecite{Lewtas-PRB-2010}\\
\end{tabular}
\end{ruledtabular}
\end{table*}

\begin{figure} [htp!]
\centering\includegraphics[width=3.4in]{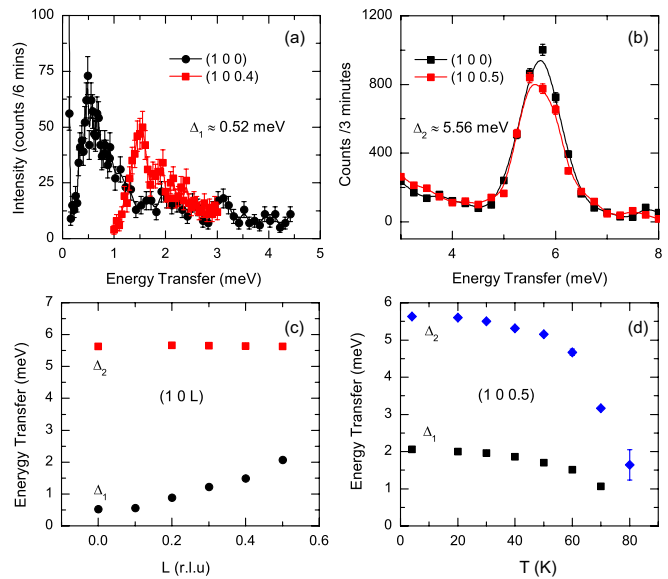}
\caption{\label{fig:tasdata}(Color online) Magnetic excitations in
Y$_{0.7}$Lu$_{0.3}$MnO$_3$ measured using triple-axis spectrometers.
(a) The $\Delta_1$ $\approx$ 0.52 meV gap and its q-dependence along
$L$  comparing constant wave-vector scans measured at (1 0 0) and (1
0 0.4) using the CG-4C cold neutron TAS at T = 4 K. (b) The
$\Delta_{2}$ $\approx$ 5.65 meV gap and its q-dependence along $L$
comparing constant wave-vector scans measured at (1 0 0) and (1 0
0.5) at T = 4 K using the HB-1A thermal neutron TAS. (c) Spin wave
dispersion along the L-direction measured at 4 K. (d) Temperature
evolution of the magnetic excitations measured at (1 0 0.5).}
\end{figure}

A summary of the TAS data along the $c$-axis is shown in
Fig.~\ref{fig:tasdata}. As illustrated in Fig.~\ref{fig:tasdata}
(a), $\Delta_{1}$ $\approx$ 0.52 meV at (1 0 0) and the excitation
shifts to $\sim$ 1.5 meV at (1 0 0.4). On the other hand, the
$\Delta_{2}$ $\approx$ 5.56 meV excitation shows no $L$ dependence
as depicted in Fig.~\ref{fig:tasdata} (b). Fig.~\ref{fig:tasdata}
(c) plots the dispersion curves along the $L$ direction for both
excitations constructed from a series of energy scans at constant
wave-vector. The data points were determined by fitting the energy
scans at constant wave-vector assuming Gaussian peak-shapes. The
0.52 meV gap shows modest dispersion along the $L$-direction with a
maximum energy shift of $\sim$ 1.5 meV and the 5.56 meV gap is
dispersionless along $L$ within the instrumental resolution. The TAS
data points along $H$ are plotted in Fig.~\ref{fig:spinwave} to show
a good agreement with the HYSPEC data. Weak dispersions were
observed along $L$ out of the hexagonal plane and strong dispersions
were observed along $H$ in the hexagonal plane consistent with the
layered magnetic structure of Y$_{1-y}$Lu$_{y}$MnO$_3$.
Fig.~\ref{fig:tasdata} (d) shows the temperature evolution of the
excitations measured at (1 0 0.5). The energy of both excitations
decrease with increasing temperature and vanish at $T_N$ confirming
their magnetic origin. In summary, the q and temperature dependence
of the magnetic spectra of Y$_{0.7}$Lu$_{0.3}$MnO$_3$ show very
similar behavior comparing to YMnO$_3$ and LuMnO$_3$ except for the
significantly smaller $\Delta_{1}$ $\approx$ 0.52 meV.

\begin{figure} [htp!]
\centering\includegraphics[width=2.6in]{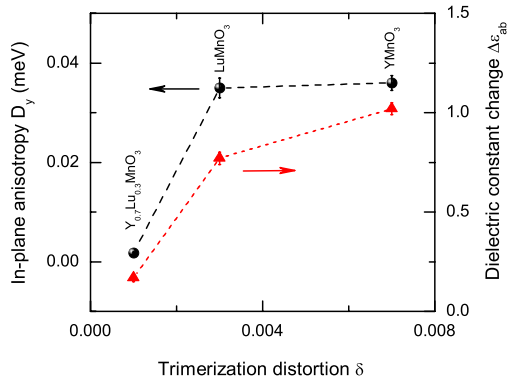}
\caption{\label{fig:delta}(Color online) In-plane single ion
anisotropy $D_y$ versus trimerization distortion $\delta$ (ball,
left axis), and critical dielectric constant change
$\Delta$$\varepsilon_{ab}$ versus $\delta$ (triangle, right axis)
for Y$_{0.7}$Lu$_{0.3}$MnO$_3$, YMnO$_3$, and LuMnO$_3$.}
\end{figure}

In order to make a quantitative comparison between YMnO$_3$,
Y$_{0.7}$Lu$_{0.3}$MnO$_3$,  and LuMnO$_3$, we model the dispersion
of the magnetic excitations in Fig.~\ref{fig:spinwave} using the
same linear spin wave analysis that has been applied to analyze the
HoMnO$_3$ and YMnO$_3$ INS data
\cite{Vajk-PRL-2005,YMnO3-INS1-cryst}. The model, defined by the
Hamiltonian ${\mathcal H}$ = J $\sum_{<i,j>}{\mathbf S}_{i}\cdot
{\mathbf S}_{j}$ + $D_{y} \sum_{i}({\mathbf S}_{i}^{y})^2 + D_{z}
\sum_{i}({\mathbf S}_{i}^{z})^2$, takes into account the
nearest-neighbor exchange interaction $J$, and two single ion
anisotropy terms $D_y$ and $D_z$ representing the in-plane and
out-of-plane anisotropy constants, respectively. Three modes are
obtained with the following dispersion equation:
\begin{equation}
{\mathcal {\hbar\omega}_i} = 3JS
\sqrt{[(1+4z_{i}+2D_{z}/{3J})(1-2z_{i}+2D_{y}/{3J})]},
\label{eq:dispersion}
\end{equation}
where $z_i$ ($i$ = 1, 2 or 3) is defined by lattice Fourier sums
identical to Ref. \onlinecite{Vajk-PRL-2005}, and $S$ = 2 for
Mn$^{3+}$ ions. A fit to the data using Eq.~\ref{eq:dispersion}
yields $J$ = 2.57 (0.05) meV, $D_y$ = 0.0017 (0.0002) meV, and $D_z$
= 0.35 (0.01) meV. Our INS data can be well described by the model
as depicted in Fig.~\ref{fig:spinwave} comparing the three spin wave
modes calculated using the fitting parameters to the measured
magnetic spectra. The obtained fitting parameters are listed in
Table \ref{tb:comp} in comparison with the results for YMnO$_3$ and
LuMnO$_3$ \cite{note2}. The lattice constants, $T_N$, $J$ and $D_z$
values of Y$_{0.7}$Lu$_{0.3}$MnO$_{3}$ all fall in between the
values of YMnO$_3$ and LuMnO$_3$. This is consistent with the
lattice parameters and unit cell volume being contracted from
YMnO$_3$ to LuMnO$_3$. However, a much smaller $D_y$ value was
obtained for Y$_{0.7}$Lu$_{0.3}$MnO$_{3}$.

At the zone center, the energy of the two magnetic excitations can
be described as $\Delta_{1}$ = $3JS[(2D_y/3J)(3)]^{1/2}$ and
$\Delta_{2}$ = $3JS[(2D_z/3J)(3/2)]^{1/2}$ corresponding to the
in-plane and out-of-plane single ion anisotropy gaps
\cite{YMnO3-INS1-cryst}. Unlike YMnO$_3$ and LuMnO$_3$ which have
almost the same value of $\Delta_1$ $\sim$ 2.5 meV regardless the
opposite trimerization distortion direction in these materials, a
smaller in-plane anisotropy gap $\Delta_1$ $\sim$ 0.52 meV was
observed in Y$_{0.7}$Lu$_{0.3}$MnO$_{3}$ that yields a significantly
smaller $D_y$ $\sim$ 0.0017 meV value. Previous systematic study
indicates that the Mn atomic position in
Y$_{0.7}$Lu$_{0.3}$MnO$_{3}$ is very close to the critical value ($x
\sim$ 1/3) \cite{Park-PRB-2010}, the Mn trimerization distortion in
Y$_{0.7}$Lu$_{0.3}$MnO$_{3}$ is much weaker comparing to YMnO$_3$
and LuMnO$_3$. The in-plane single ion anisotropy $D_y$ is coupled
to the Mn trimerization in the hexagonal plane thus is very
sensitive to the strength of trimerization distortion
\cite{Solovyev-PRB-2012-theory}. We attribute the observed small
in-plane anisotropy gap to the weaker Mn trimerization distortion in
Y$_{0.7}$Lu$_{0.3}$MnO$_{3}$. It is thus of great interest to see
how the weaker Mn trimerization distortion affects the magnitude of
the ME coupling. If the ME effect is directly linked to the Mn
trimerization distortion, we would expect a much weaker in-plane
dielectric anomaly in Y$_{0.7}$Lu$_{0.3}$MnO$_{3}$ which is indeed
what we observed in the dielectric constant measurements. As
illustrated in Fig.~\ref{fig:tn} (c), at $T_N$, no anomaly was
observed in $\varepsilon_{c}$ consistent with previous reports,
whereas a weaker dielectric anomaly was observed in
$\varepsilon_{ab}$. Fig.~\ref{fig:tn} (d) compares the in-plane
dielectric constant $\varepsilon_{ab}$ between YMnO$_3$, LuMnO$_3$
and Y$_{0.7}$Lu$_{0.3}$MnO$_{3}$ ($\varepsilon_{ab}$ values for
YMnO$_3$ and LuMnO$_3$ are taken from Ref.
\onlinecite{Katsufuji-2001} and plotted in Fig.~\ref{fig:tn} (d)
with a -2.1 and 0.3 offset, respectively) and it clearly shows that
the dielectric anomaly observed in Y$_{0.7}$Lu$_{0.3}$MnO$_{3}$ is
much weaker comparing to YMnO$_3$ and LuMnO$_3$.

Table \ref{tb:comp} compares our Y$_{0.7}$Lu$_{0.3}$MnO$_{3}$
results to the ones reported for YMnO$_3$ and LuMnO$_3$ from the
previous study. We define a trimerization distortion parameter
$\delta$ to reflect the strength of the trimerization distortion,
$\delta$ = $\mid x-x_c\mid$ as depicted in
Fig.~\ref{fig:trimerization}. The $\delta$ values listed in Table
\ref{tb:comp} are based on the data reported in Ref.
\onlinecite{Park-PRB-2010}. We also define a critical dielectric
constant change parameter $\Delta$$\varepsilon_{ab}$ to represent
the magnitude of the ME coupling at $T_{N}$. As shown in
Fig.~\ref{fig:tn} (c), the $T < T_N$ and $T
> T_N$ $\varepsilon_{ab}$ data are fit to a linear function
respectively, and $\Delta$$\varepsilon_{ab}$ is the difference
between these two fittings at $T_N$. The $D_y$ vs. $\delta$ and
$\Delta$$\varepsilon_{ab}$ vs. $\delta$ for YMnO$_3$,
Y$_{0.7}$Lu$_{0.3}$MnO$_{3}$, and LuMnO$_3$ are plotted in
Fig.~\ref{fig:delta}. It shows that both $D_y$ and
$\Delta$$\varepsilon_{ab}$ decrease with decreasing $\delta$
indicating strong correlations between the strength of trimerization
distortion and the magnitude of ME coupling.

In summary, our INS study of Y$_{0.7}$Lu$_{0.3}$MnO$_{3}$ reveals a
small in-plane single ion anisotropy gap that coincides with a
weaker dielectric anomaly in $\varepsilon_{ab}$.  This is attributed
to a much weaker Mn trimerization distortion in
Y$_{0.7}$Lu$_{0.7}$MnO$_{3}$ due to the doping influence of the Mn
atomic position $x$ $\sim$ 1/3. These results provide strong
evidence that the Mn trimerization is responsible for the ME effect
in Y$_{1-y}$Lu$_{y}$MnO$_{3}$ and the magnitude of ME coupling is
coupled to the strength of the trimerization distortion. High
resolution neutron diffraction study have shown that the Mn
trimerization is a systematic feature in RMnO$_3$
\cite{Fabreges-PRL-2009}, this finding may shed light on a deeper
understanding of the multiferroic phenomenon in this series of
materials inviting further theoretical investigations.

We acknowledge valuable discussions with Randy Fishman. Work at the
High Flux Isotope Reactor and Spallation Neutron Source, Oak Ridge
National Laboratory, was sponsored by the Scientific User Facilities
Division, Office of Basic Energy Sciences, U.S. Department of
Energy. The single crystal growth and neutron scattering
work at UTK/Rice is supported by the U.S. DOE
BES under Grant No. DE-FG02-05ER46202.

\end{document}